\documentclass[final,3p,times,twocolumn,longtitle, aastex6, sort&compress]{elsarticle}

\usepackage{lineno}
\usepackage{graphicx}
\usepackage{color}
\usepackage{dcolumn}
\usepackage{bm}
\usepackage{multirow,array,tabularx}
\usepackage{natbib}
\usepackage{hyperref}

\usepackage{amssymb}
\usepackage{amsthm}
\usepackage{mathtools}

\newcommand*{\Rice}{Rice University, Houston, Texas 77005}
\newcommand*{\FIU}{Florida International University, Miami, Florida 33199}
\newcommand*{\ANL}{Argonne National Laboratory, Argonne, Illinois 60439}

\newcommand*{\ASU}{Arizona State University, Tempe, Arizona 85287-1504}

\newcommand*{\CSUDH}{California State University, Dominguez Hills, Carson, CA 90747}

\newcommand*{\CANISIUS}{Canisius College, Buffalo, NY}

\newcommand*{\CMU}{Carnegie Mellon University, Pittsburgh, Pennsylvania 15213}

\newcommand*{\CUA}{Catholic University of America, Washington, D.C. 20064}

\newcommand*{\SACLAY}{IRFU, CEA, Universit'e Paris-Saclay, F-91191 Gif-sur-Yvette, France}

\newcommand*{\CNU}{Christopher Newport University, Newport News, Virginia 23606}

\newcommand*{\UCONN}{University of Connecticut, Storrs, Connecticut 06269}

\newcommand*{\DUKE}{Duke University, Durham, North Carolina 27708-0305}

\newcommand*{\FU}{Fairfield University, Fairfield CT 06824}

\newcommand*{\FERRARAU}{Universita' di Ferrara , 44121 Ferrara, Italy}

\newcommand*{\FSU}{Florida State University, Tallahassee, Florida 32306}

\newcommand*{\Genova}{Universit$\grave{a}$ di Genova, 16146 Genova, Italy}

\newcommand*{\GWUI}{The George Washington University, Washington, DC 20052}

\newcommand*{\ISU}{Idaho State University, Pocatello, Idaho 83209}

\newcommand*{\INFNFE}{INFN, Sezione di Ferrara, 44100 Ferrara, Italy}

\newcommand*{\INFNGE}{INFN, Sezione di Genova, 16146 Genova, Italy}

\newcommand*{\INFNRO}{INFN, Sezione di Roma Tor Vergata, 00133 Rome, Italy}

\newcommand*{\INFNTUR}{INFN, Sezione di Torino, 10125 Torino, Italy}

\newcommand*{\ORSAY}{Institut de Physique Nucl\'eaire, CNRS/IN2P3 and Universit\'e Paris Sud, Orsay, France}

\newcommand*{\Juelich}{Institute fur Kernphysik (Juelich), Juelich, Germany}

\newcommand*{\ITEP}{Institute of Theoretical and Experimental Physics, Moscow, 117259, Russia}

\newcommand*{\JMU}{James Madison University, Harrisonburg, Virginia 22807}

\newcommand*{\KNU}{Kyungpook National University, Daegu 41566, Republic of Korea}

\newcommand*{\MISS}{Mississippi State University, Mississippi State, MS 39762-5167}

\newcommand*{\UNH}{University of New Hampshire, Durham, New Hampshire 03824-3568}

\newcommand*{\NSU}{Norfolk State University, Norfolk, Virginia 23504}

\newcommand*{\OHIOU}{Ohio University, Athens, Ohio  45701}

\newcommand*{\ODU}{Old Dominion University, Norfolk, Virginia 23529}

\newcommand*{\ROMAII}{Universita' di Roma Tor Vergata, 00133 Rome Italy}

\newcommand*{\MSU}{Skobeltsyn Institute of Nuclear Physics, Lomonosov Moscow State University, 119234 Moscow, Russia}

\newcommand*{\SCAROLINA}{University of South Carolina, Columbia, South Carolina 29208}

\newcommand*{\TEMPLE}{Temple University,  Philadelphia, PA 19122 }

\newcommand*{\JLAB}{Thomas Jefferson National Accelerator Facility, Newport News, Virginia 23606}

\newcommand*{\UTFSM}{Universidad T\'{e}cnica Federico Santa Mar\'{i}a, Casilla 110-V Valpara\'{i}so, Chile}

\newcommand*{\EDINBURGH}{Edinburgh University, Edinburgh EH9 3JZ, United Kingdom}

\newcommand*{\GLASGOW}{University of Glasgow, Glasgow G12 8QQ, United Kingdom}

\newcommand*{\VIRGINIA}{University of Virginia, Charlottesville, Virginia 22901}

\newcommand*{\VCU}{Virginia Commonwealth University, Richmond, VA 23220}

\newcommand*{\WM}{College of William and Mary, Williamsburg, Virginia 23187-8795}

\newcommand*{\YEREVAN}{Yerevan Physics Institute, 375036 Yerevan, Armenia}

\journal{ Physics Letters B}

\begin{document}

\begin{frontmatter}

\title{First Measurement of $\Xi^-$ Polarization in Photoproduction}

\author[toFIU,toRice]{J.~Bono\fnref{toFermi}}\ead{jbono@fnal.gov}
\author[toFIU]{L.~Guo}\ead{leguo@fiu.edu}
\author[toFIU]{B.A.~Raue}
\author[toFIU]{S. Adhikari}
\author[toJuelich]{M.C.~Kunkel}
\author[toODU]{K.P. ~Adhikari\fnref{toNOWMISS}}
\author[toFSU]{Z.~Akbar}
\author[toODU]{M.J.~Amaryan}
\author[toSACLAY]{J.~Ball}
\author[toINFNFE]{L. Barion}
% ...

\fntext[toFermi]{Current address: Particle Physics Division, Fermi National Accelerator Laboratory, Batavia, IL 60510} 
\fntext[toNOWMISS]{Current address: Mississippi State University, MS 39762-5167 }
 \fntext[toNOWIABFU]{Current address: Imam Abdulrahman Bin Faisal University,  Industrial Jubail 31961, Saudi Arabia}
 \fntext[toNOWUK]{Current address: LEXINGTON, KENTUCKY 40506 }
 \fntext[toNOWISU]{Current address: Pocatello, Idaho 83209 }
 \fntext[toNOWINFNGE]{Current address: 16146 Genova, Italy }

 \address[toFIU]{\FIU} 
 \address[toRice]{\Rice}
 \address[toJuelich]{\Juelich} 
% ...
\address[toSACLAY]{\SACLAY} 
% ...
\address[toFSU]{\FSU} 
% ...
\address[toINFNFE]{\INFNFE} 
% ...
\address[toODU]{\ODU} 
% ...

  
\date{\today}

\begin{abstract}

Despite decades of studies of the photoproduction of hyperons, both their production mechanisms and their spectra of excited states are still largely unknown.  While the parity-violating weak decay of hyperons offers a means of measuring their polarization, which could help discern their production mechanisms and identify their excitation spectra, no such study has been possible for doubly strange baryons in photoproduction, due to low production cross sections. However, by making use of the reaction  $\gamma p \to K^+ K^+ \Xi^-$,  we have measured, for the first time, the induced polarization, $P$, and the transferred polarization from circularly polarized real photons, characterized by $C_x$ and $C_z$, to recoiling $\Xi^-$s. The data were obtained using the CEBAF Large Acceptance Spectrometer (CLAS) at Jefferson Lab for photon energies from just over threshold (2.4 GeV) to 5.45 GeV. These first-time measurements are compared, and are shown to broadly agree, with model predictions in which cascade photoproduction proceeds through the decay of intermediate hyperon resonances that are produced via relativistic meson exchange, offering a new step forward in the understanding of the production and polarization of doubly-strange baryons.

\end{abstract}


\end{frontmatter}




\section{Introduction}

The polarization of hyperons can be measured through the angular distribution of their parity-violating weak decay products, providing insight into the mechanisms behind their production. Such measurements involving the photo- and electroproduction of Strangeness number $S=-1$ 
hyperons~\cite{Paterson:2016,Tran:1998qw,Carman:2002se,Glander:2003jw,Zegers:2003ux, McNabb:2003nf, Sumihama:2005er, Bradford:2006ba, McCracken:2009ra, Dey:2010hh,  
Carman:2009fi, Gabrielyan:2014zun} have led to significant progress in understanding the excitation spectrum of $S=0$ 
nucleons~\cite{Anisovich:2007bq,Nikonov:2007br, Anisovich:2010an, Anisovich:2011ye, Anisovich:2011su, delaPuente:2008bw, Maxwell:2012tn, Maxwell:2014kba, Maxwell:2015psa, 
Maxwell:2016hdx, Penner, Doring, Kamano}.  A similar opportunity exists in studying the 
polarization of $S=-2$ cascades, which could prove vital for understanding their production mechanism and in gaining an understanding of the excitation spectrum of $S=-1$ hyperons. 
However, because of the cascade's low production cross section and the resulting lack of available data, no previous cascade polarization measurements exist in either photo- or 
electroproduction.

\begin{figure}[tp] 
\includegraphics[width=0.45\textwidth]{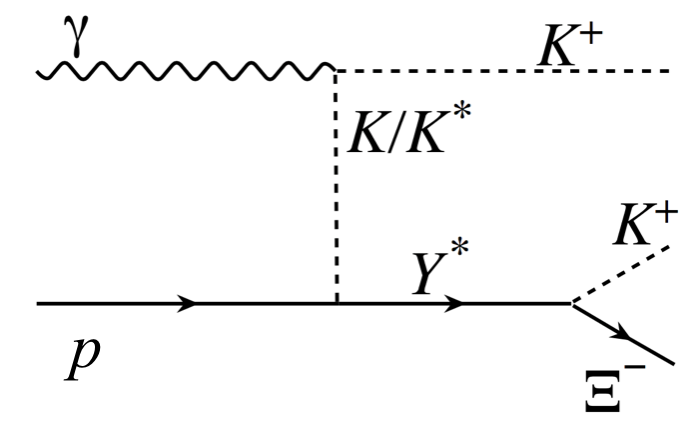}
\caption{A possible Feynman diagram of $\Xi^-$ photoproduction via the decay of intermediate hyperon resonances in $t$-channel $K/K^*$ exchange, which is a major component in the production models of Nakayama~\cite{Nakayama:2006ty,Man:2011np}. } 
\label{fig:reaction}
\end{figure}

The CLAS collaboration has reported cross-section measurements for cascade photoproduction~\cite{Price:2004xm, Guo:2007dw}. In these data, a strong back-angle 
peaking in the center-of-momentum cascade angular distribution ($\cos\theta_\Xi$) was observed, which along with the invariant mass distributions of the $K^+\Xi^-$ system, suggested the significant role that intermediate hyperon resonances with masses of about 2 GeV play in cascade photoproduction. These results generated theoretical interest in understanding the 
production mechanism behind $S=-2$ states. In particular, Refs.~\cite{Nakayama:2006ty,Man:2011np} found it is necessary to include the contributions from the decay of high-mass hyperons (up to $\Lambda$(1890)) that are predominately produced in $t$-channel $K/K^*$ exchange, as illustrated in Fig.~\ref{fig:reaction}, to explain the CLAS cross-section measurements~\cite{Guo:2007dw}. Furthermore, Ref.~\cite{Man:2011np} investigated the role of the addition of high-spin hyperon states around 2~GeV and found significant contributions from spin/parity $J^P = \frac{5}{2}^{\pm}$ and $\frac{7}{2}^{\pm}$ resonances. In particular,  the inclusion of the  $\Sigma(2030) \frac{7}{2}^+$ state improved the model's agreement with the data. 

These earlier photoproduction data from CLAS did not have either beam or target polarization, and no study on induced polarization was carried out.  But as pointed out in 
Ref.~\cite{Man:2011np}, both the induced and transferred polarization of the cascade ground state are sensitive to the production mechanism, particularly, the mass, spin and 
parity of intermediate hyperon resonances, as well as to the mesonic exchange mechanisms.

The majority of early data for hyperon and cascade spectroscopy was generated using $K^-$ beams on nuclear targets. 
However, the significance of the $Y^* \rightarrow K \Xi$ decay has never been firmly established except for the small branching ratios and branching-ratio upper limits reported for 
$\Lambda(2100) \frac{7}{2}^-$ and  $\Sigma(2030) \frac{7}{2}^+$~\cite{Litchfield:1971, Muller:1969, Tripp:1967, Burgun:1968} in the 1960's and 1970's. In general, the excitation spectrum for $S=-1$ hyperons also remains under-explored, particularly in the high 
mass ($>2$~GeV) region.   When compared with model predictions, cascade polarization measurements can build on the evidence for or against intermediate hyperon resonances as 
the dominant production mode, discriminate among the candidate exchange mechanisms, and even point to the existence of higher mass/spin hyperons.

The understanding of the ground state cascade production mechanism is not limited to its connection to the intermediate hyperon resonances. The current spectrum of experimentally established excited cascade states has remained virtually unchanged in the past thirty years~\cite{Olive:2016xmw}. At present, just six states are 
considered to have solid experimental evidence, and only half of these have established spin and parity.  Furthermore, the number of cascade (as well as hyperon) states that 
appear in the most recent lattice QCD calculations ~\cite{Edwards:2013} are nearly as numerous as predicted by early constituent quark models~\cite{Capstick:1986}.  
Understanding the production of excited cascades cannot be fully achieved without a better understanding of the ground state production, including polarization measurements. This manuscript reports the first measurements of both induced and transferred polarization of cascade baryons in photoproduction.


\section{Experimental Details}

A large-statistics dataset with an integrated luminosity of $68$ pb$^{-1}$ was collected with CLAS~\cite{Mecking:2003zu} using a circularly polarized, tagged photon 
beam~\cite{Sober:2000we} of energy range $1.1$ to $5.4$~GeV incident on a liquid hydrogen target~\cite{g12note2}. The photon beam was produced from a longitudinally polarized primary electron beam of energy $5.7$~GeV, incident on a gold radiator. The electron-beam's helicity was flipped pseudo-randomly at a rate of $30$~Hz and was measured periodically by a M{\o}ller polarimeter, yielding a degree of polarization of $0.68,$ averaged over the entire run period.  The degree of circular photon polarization was calculated and is known to be proportional to the electron beam polarization, and to increase as a function of the ratio of photon energy to the energy of the primary electron beam~\cite{Olsen:1959zz}. 
The target consisted of a 40-cm-long cylindrical cell containing liquid hydrogen. Momentum  information for charged particles were obtained via tracking through three regions of multiwire drift chambers~\cite{Mestayer:2000we}, with the region-two drift chambers inside a toroidal magnetic field that was generated by six superconducting coils. 
Scintillators~\cite{Smith:1999ii} outside of the drift chambers were used to measure time-of-flight (TOF) information, which, when combined with the momentum 
information, provided charged-particle identification.


\begin{figure}[tb]  
\includegraphics[width=0.47\textwidth]{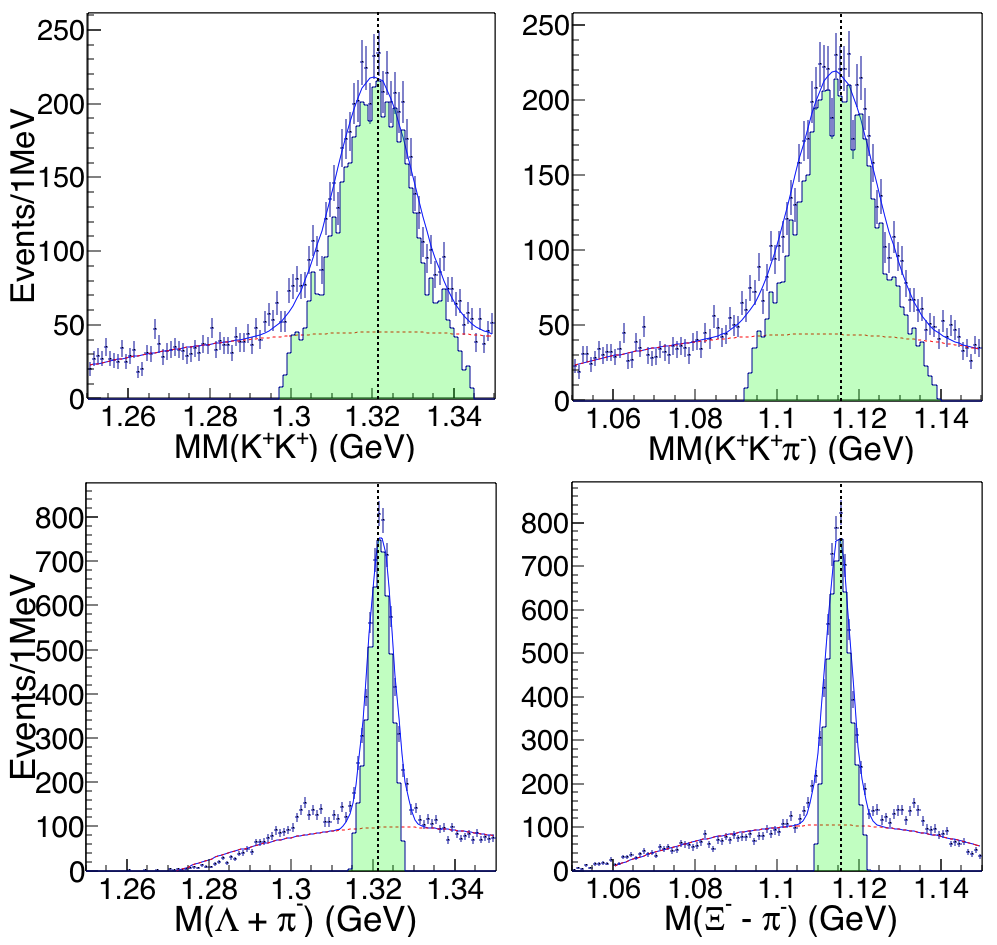} 
\caption{Mass distributions for all events passing cuts on timing, detected particle mass, and vertex location are shown by the data points with error bars. Top left: Missing mass spectrum of the $K^+ K^+$ system;  Top right: Missing mass spectrum of the $K^+ K^+ \pi^-$ system; Bottom left: Invariant mass spectrum of the $\Lambda \pi^-$ system; Bottom right: Invariant mass spectrum as reconstructed from the four-momentum difference of the $\Xi^-$ and $\pi^-$ system.  In all plots, a Gaussian is fit to the signal over a polynomial background (dashed red line).  The same distributions after applying the hypersphere cuts are shown by the filled histograms.	The vertical lines represents the known $\Lambda$ or $\Xi^-$ masses. Detection of the $\pi^-$ originating the $\Lambda$ decay, rather than the $\Xi^-$ decay, is evident in the left and right of the signal region, in the bottom left and bottom right plots, respectively.}
\label{fig:signal}
\end{figure}

\section{Analysis}

Initial event selection required timing coincidences between the photon tagger and the passage of two charged particles through the CLAS detector.  The photons that produced the event were selected using vertex information obtained from tracking, and the timing information from a start counter~\cite{Sharabian:2005kq}, which surrounded the target.  The time that an event occurred at its vertex, as measured by the start counter, was required to be within $\pm 1$~ns of the photon time provided by the accelerator radio-frequency signal. Furthermore, the vertex time determined from the TOF system was required to be within $\pm 1$~ns of the photon time for all detected charged particles.

The next step in the identification of the $\gamma p\rightarrow K^+K^+\Xi^-$ reaction with the subsequent decay of $\Xi^- \rightarrow \Lambda \pi^-$ was selecting events with  three charged mesons, $K^+$, $K^+$, and $\pi^-$,  detected.  Their momentum was corrected for the energy loss in the target region, as well as other detector effects such as misalignments and errors in the magnetic field map. The signals were then extracted using the following four mass distributions:

\begin{enumerate}
\item Missing mass in the $\gamma  p  \to K^+  K^+ (X)$ reaction, where $X$ indicates the missing particle, labeled as $MM(K^+K^+)$.
 \item Missing mass in the $\gamma  p \to  K^+  K^+  \pi^- (X)$ reaction, where $X$ indicates the missing particle, labeled as $MM(K^+K^+\pi^-)$.
 \item Invariant mass of the $(\Lambda+\pi^-)$ system, labeled as $M(\Lambda+\pi^- )$, and where the known $\Lambda$ mass, 1115.683 GeV~\cite{PDG}, was combined with the missing three-momentum of the $K^+K^+\pi^-$ system to define the $\Lambda$ four-momentum vector.

 \item Invariant mass reconstructed from the four-momentum difference of the $\Xi^-$  and $\pi^-$ system, labeled as $M(\Xi^--\pi^-)$, and where the known $\Xi^-$ mass, 1321.71 GeV~\cite{PDG}, was combined with the missing three-momentum of the $K^+K^+$ system to define the $\Xi^-$ four-momentum vector.

\end{enumerate}
The mass distributions for events passing cuts on event timing, event vertex location, and detected particle mass are shown by the data points with error bars in Fig.~\ref{fig:signal}.  
Clear signals for the  $\Lambda$ and $\Xi^-$ are seen. 

Instead of cutting on individual mass distributions, each of the above quantities was scaled by the reciprocal of their individually associated 3$\sigma$ width, and treated as orthogonal displacements in a four dimensional space. A composite cut was then placed on the volume of the hypersphere that was constructed from the scaled displacements. The width $\sigma$, of each mass distribution was measured by fitting it with a Gaussian plus a polynomial to model the signal and background, as shown by the fits in 
Fig.~\ref{fig:signal}. The hypersphere coordinates were defined as,
\begin{eqnarray}
	x_{1} =  \left[MM\left(K^{+}K^{+}\right) - \Xi_{mass}^{-}\right]/3\sigma_1, \\
	x_{2} = \left[MM\left(K^{+} K^{+}\pi^{-}\right) - \Lambda_{mass}\right]/3\sigma_2, \nonumber \\
	x_{3} = \left[M\left(\Lambda+\pi^{-}\right)   - \Xi_{mass}^{-}\right]/3\sigma_3, \nonumber \\
	x_{4} = \left[M\left(\Xi^{-}-\pi^{-}\right) - \Lambda_{mass}\right]/3\sigma_4, \nonumber \\
	r = \sqrt{ x_{1}^{2}  +  x_{2}^{2} +  x_{3}^{2}  +  x_{4}^{2} }, \nonumber
\end{eqnarray}
where $\sigma_n$ denotes the Gaussian width of the associated quantity as displayed in Fig.~\ref{fig:signal}. A cut on the hypersphere radius $r$ represents a simultaneous cut on all four mass quantities, where a $3\sigma$ cut corresponds to taking events within the hypervolume defined by $r < 1$. This cut, as opposed to simply rectangular cuts on the masses, allowed the best signal to background ratio, even though $x_i$'s are not totally independent. The final data sample of  $5143$ events are shown in the filled histograms in Fig.~\ref{fig:signal}.


\begin{figure}[tb] 
\begin{center}
\includegraphics[width=0.4\textwidth]{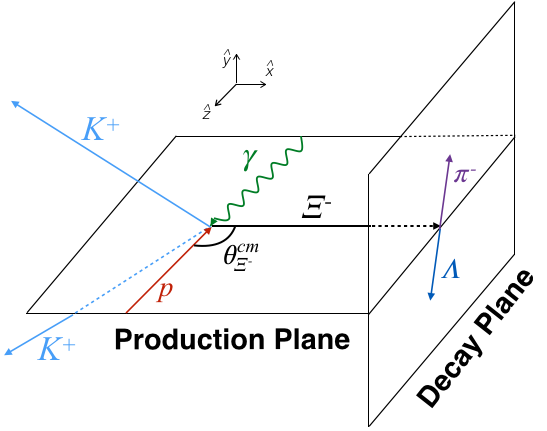}
	\caption{Plane and angle definitions for the polarization observables of $C_x$, $C_z$, and $P$. See the text for a full description of the coordinates.} 
\label{fig:plane}
\end{center}
\end{figure} 

The $\Xi^-$ polarization is related to the angular distribution of the decay $\pi^-$ as measured in the rest frame of the $\Xi^-$ by~\cite{Huang}
\begin{equation}
	I(\cos\theta^{\hat{i}}_\pi)  =  \frac{N}{2}(1 - P_{\Xi_i} \alpha \cos \theta^{\hat{i}}_{\pi} ),
\end{equation}
where $\theta^{\hat{i}}_\pi$ is the pion angle relative to the $i=x,y,$ or $z$ axes in the $\Xi^-$ rest frame, $N$ is the total number of events in the 
$I(\cos\theta^{\hat{i}}_\pi)$ distribution, $P_{\Xi_i}$ is the $i$-component of the $\Xi^-$ polarization, and $\alpha$ is the $\Xi^-$ weak-decay asymmetry 
or analyzing power with $\alpha = -0.458\pm 0.012$ ~\cite{Olive:2016xmw}. The axes are defined in the $\Xi^-$ rest-frame (Fig.~\ref{fig:plane}) as 
\begin{eqnarray}
\hat{z} = \frac{\vec{p}_{\gamma}} {|\vec{p}_{\gamma}|},\\
\hat{y} =  \frac{\hat{z} \times \vec{p}_{\Xi}}{|\hat{z} \times \vec{p}_{\Xi} |},\nonumber \\
\hat{x} = \hat{y} \times \hat{z},\nonumber
\end{eqnarray}
where $\vec{p}_\gamma$ and $\vec{p}_\Xi$ are the photon and cascade momentum vectors, respectively,
both in the center-of-momentum frame of the beam-plus-target system.
The spin observables $P$, $C_x$, and $C_z$ are connected to the recoil polarization $\vec{P}_{\Xi}$ through,
\begin{equation}
\label{eqn:PP}
     \begin{aligned}\
       & P_{\Xi_x} = P_\odot C_x, \\
       & P_{\Xi_y} = P, \\
       & P_{\Xi_z} = P_\odot C_z,
     \end{aligned}
\end{equation}
where $P_\odot$ is the degree of photon-beam polarization.

The induced polarization, $P$, can be extracted from the forward-backward asymmetry, $A_y$, of the pion angular distribution. This method has 
the advantage of the cancelation of detector-acceptance effects, which follows from the fact that the polarization axis $\hat{y}$ points isotropically in the lab 
frame. The asymmetry is defined as,
\begin{equation}
\label{eq:Ay}
	A_y \equiv \frac{N^+_y - N^-_y}{N^+_y + N^-_y},
\end{equation}
where  $N^+_y$ and $N^-_y$ represent the number of events with $\cos\theta_\pi^y$ as positive and negative, respectively. The asymmetry is related to the induced $\Xi^-$ polarization by
\begin{equation}
\label{eqn:pasy}
	P = \frac {-2 A_y }{\alpha}.
\end{equation}

\begin{figure}[tb] 
\includegraphics[width=0.47\textwidth]{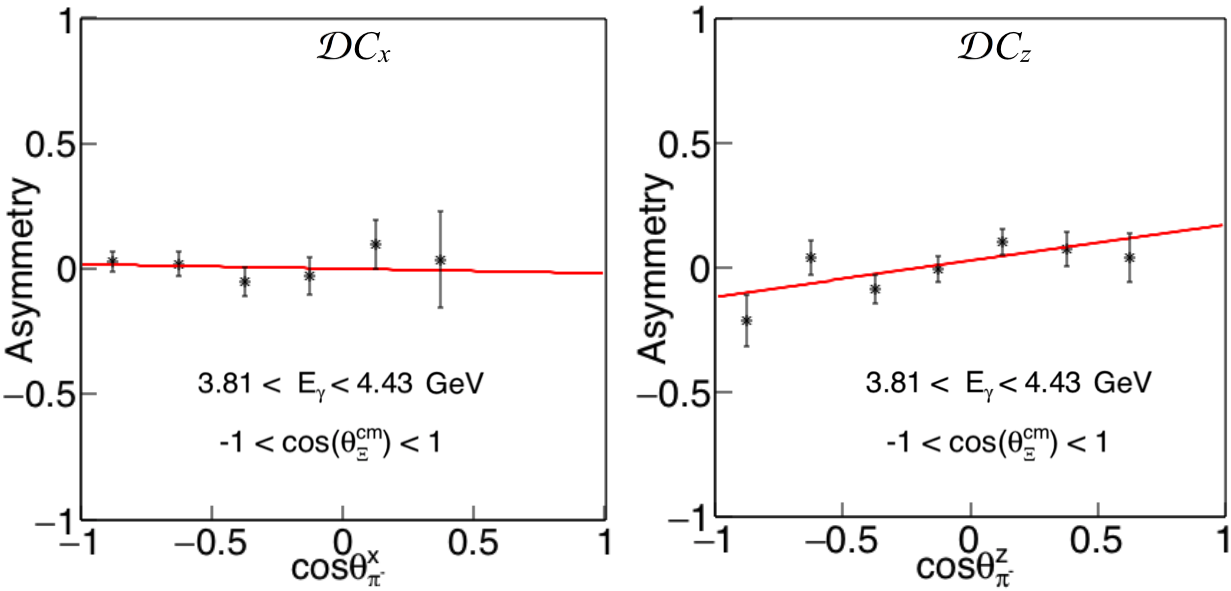}
\caption{Above shows the beam helicity asymmetries across $x$ and $z$ for the $\Xi^-$ decay, the slopes of which, along with the dilution factor, $\mathcal{D}$, are used to calculate $C_x$ and $C_z$. The events displayed include all angles between $\Xi^-$ and the $z$-axis but are limited to one photon energy bin. }
\label{fig:CxCzExample}
\end{figure}

The double polarization observables $C_x$ and $C_z$ characterize the transferred polarization of the photon to the $\Xi^-$ and are extracted from the 
photon-helicity asymmetry, 
\begin{equation}
\label{eqn:asym1}
	A = \frac{N^{+}_\text{hel} - N^{-}_\text{hel}}{N^{+}_\text{hel} + N^{-}_\text{hel}},
\end{equation}
where $N^{+}_\text{hel}$ and $N^{-}_\text{hel}$ are the number of events associated with positive and negative photon-beam helicity states, respectively. The transferred polarization is 
related to the photon-helicity asymmetry by
\begin{equation}
\label{eqn:asym2}
	\frac{-A(\cos\theta^{\hat{i}}_\pi)}{ |P_\odot| \alpha } =  C_i \cos\theta^{\hat{i}}_\pi.
\end{equation}
The value and uncertainty of $C_i$ can thus obtained from the slope of $A\cos\theta^{\hat{i}}_\pi$. Examples of the linear fits used to extract $C_x$ and $C_z$ are shown in 
Fig.~\ref{fig:CxCzExample}.  In the asymmetry defined in Equation~\ref{eqn:asym1}, systematic effects such as detector acceptance mostly cancel, since they occur irrespective of the photon helicity. 

It was found that overall around 15\% of the events surviving the final cuts were unpolarized background events. The fraction of these events were estimated in each kinematic bin by evaluating the background subtracted yield through a Gaussian fit with a polynomial background.  These events were found to have polarizations consistent with zero, thus reducing the measured polarization by the dilution factor,
\begin{equation}
\mathcal{D} = 1 - f_{\text{BG}},
\end{equation}
where $f_{\text{BG}}$ is the fraction of background events in each sample. In order to recover the true polarization, the measured polarization observables in each bin were divided by the corresponding dilution factor, the values of which were found to be between 0.82 and 0.91.

%
%
%
%

Aside from the dilution factor, three main sources of systematic uncertainty contributed to the overall uncertainties in the measurements.  
For one, systematic effects due to acceptance-related factors, including the selection of the fiducial region of the detector, were estimated by comparing the final results obtained with and without these cuts, and  were found to be, integrating over all kinematic bins, 
$\delta_{acc}P = 0.022$,  $\delta_{acc}C_x = 0.01$ and $\delta_{acc}C_z =0.052$. 
Additionally, uncertainty in the degree of photon-beam polarization, which in turn resulted from the uncertainty in the primary electron beam polarization, contributed a relative scale-type uncertainty of $\delta_{P_\cdot}C_i/C_i=0.03$.  Finally, the uncertainty in the analyzing 
power of the cascade, which is $\pm0.012$~\cite{Olive:2016xmw}, leads to a relative scale-type uncertainty of $\delta_\alpha P/P=\delta_\alpha C_i/C_i=0.026$. For both the induced and transferred polarization measurements, the statistical uncertainty dominates the cumulative systematic uncertainty.

\section{Results \& Comparison With Theory}

\begin{figure}[htb]  
\begin{center} 
\includegraphics[width=0.45\textwidth]{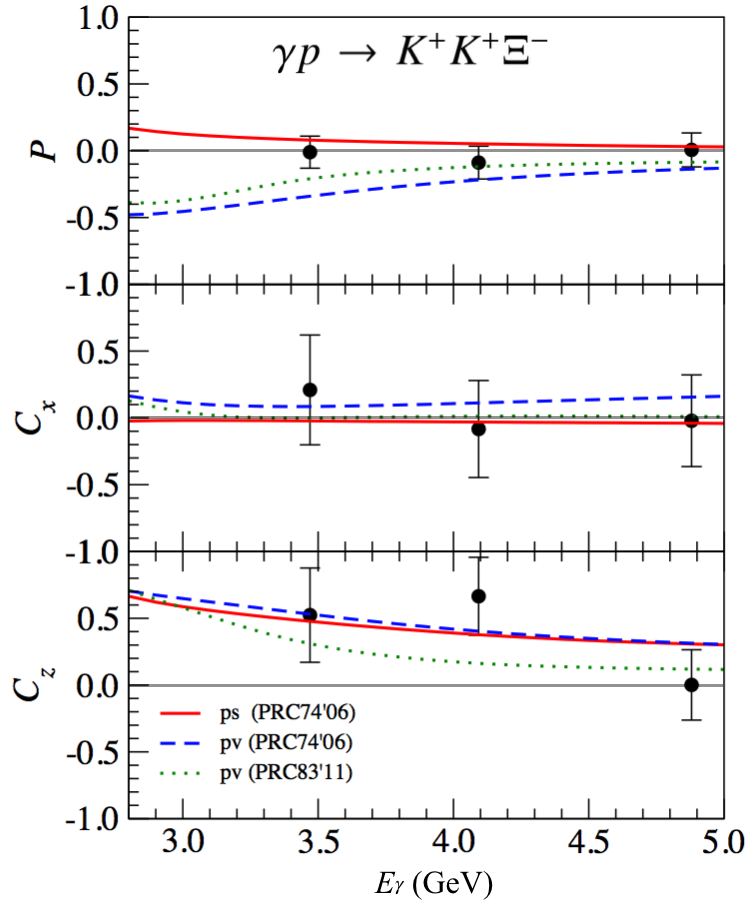} 
\caption{$P$ (top), $C_x$ (middle) and $C_z$ (bottom) as a function of $E_\gamma$ and summed over $\cos\theta_\Xi$.  The error bars represent the total uncertainty. The legend specifies pseudoscalar (ps) or pseudovector (pv) coupling, as well as the journal of publication for the associated model.}

\label{fig:PCxCzE}
\end{center} 
\end{figure}

\begin{figure}[htb]  
\begin{center} 
\includegraphics[width=0.45\textwidth]{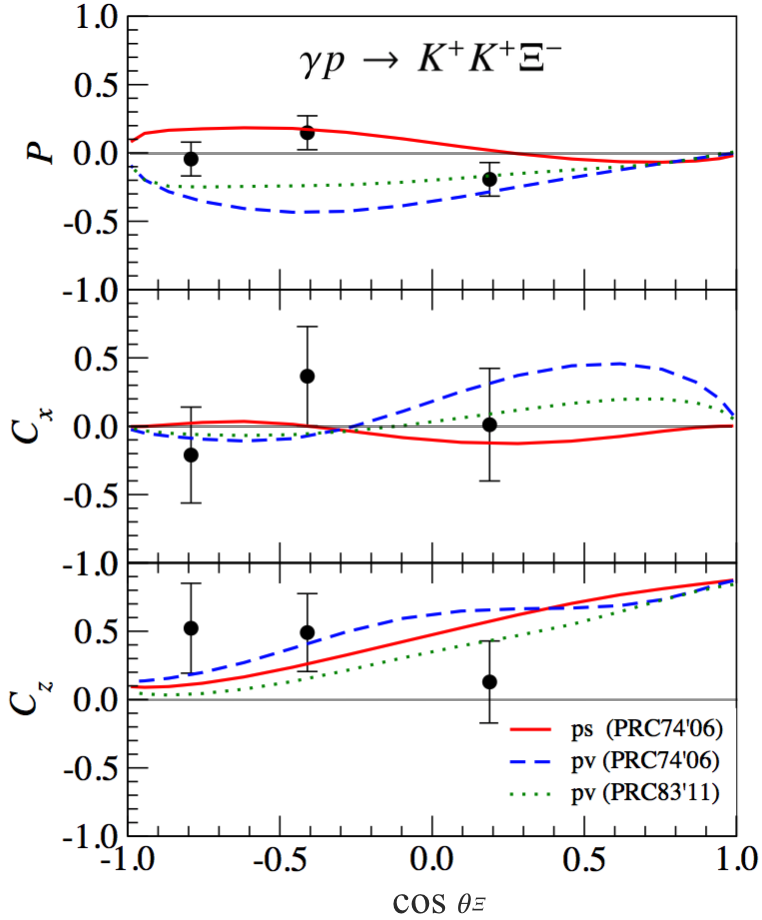} 
\caption{$P$ (top), $C_x$ (middle) and $C_z$ (bottom) as a function of $\cos\theta_\Xi$ and summed over $E_\gamma$.  Error bars and curves are the same as in 
	 Fig.~\ref{fig:PCxCzE}.}
\label{fig:PCxCzXi}
\end{center} 
\end{figure}

\begin{figure}[htb]  
\begin{center} 
\includegraphics[width=0.45\textwidth]{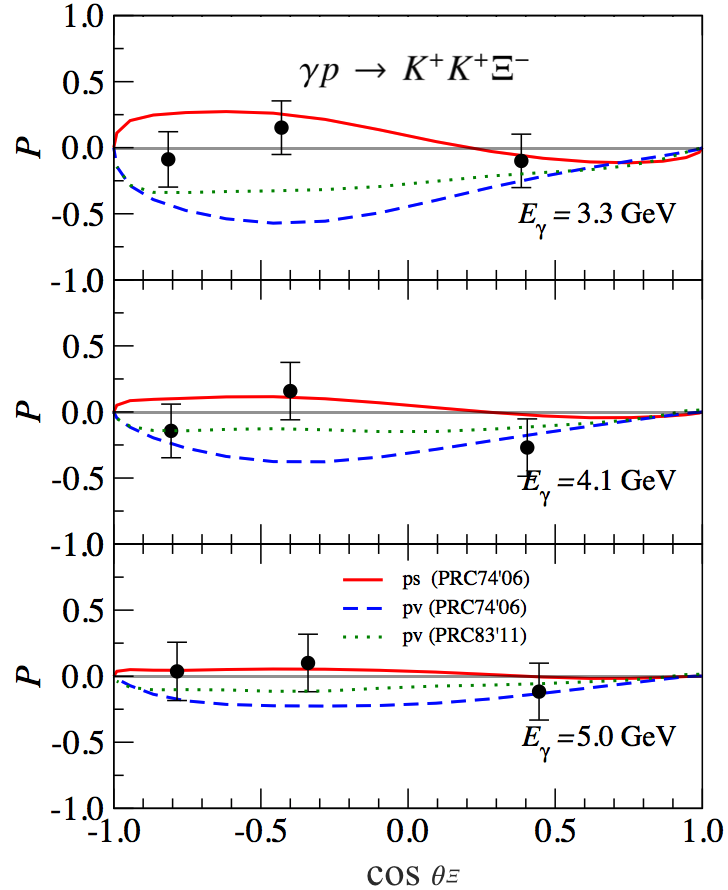} 
\caption{$P$ as a function of $\cos\theta_\Xi$ for three $E_\gamma$ bins as indicated.  Error bars and curves are the same as in 
	 Fig.~\ref{fig:PCxCzE}.}
\label{fig:P9Xi}
\end{center} 
\end{figure}

In the extraction of $P$, data were binned into nine regions defined by three bins of the cascade angle between the photon and target momenta in the c.m. frame with event-weighted average values of $\cos\theta_\Xi= - 0.79$, $-$0.41, and 0.19, and three bins of photon energy with event-weighted averages of  $E_\gamma=3.47$, 4.09, and 4.88 GeV.  Since the extractions of 
$C_x$ and $C_z$ require more events to achieve the same statistical uncertainty as $P$, these variables were binned into only three regions of $\cos\theta_\Xi$ and summed 
over $2.8\leq E_\gamma\leq 5.5$ GeV, or conversely, binned into three regions of $E_\gamma$ and summed over $-1\leq\cos\theta_\Xi\leq 1$.   The $P$ results are given in 
Table~\ref{tab:P} and the $C_x$ and $C_z$ results are given in Table~\ref{tab:Cxz}, as well as shown in Figs.~\ref{fig:PCxCzE}, \ref{fig:PCxCzXi},  and \ref{fig:P9Xi}. These results can be found in Ref.~\cite{CLASDB}. 

\begin{table}[th]
\center
\resizebox{0.49\textwidth}{!}{\begin{tabular}{c c c c c c c}
\hline\hline
$E_\gamma$ (GeV) & $\cos\theta_\Xi$ & $P$ & $\delta_{stat} P$ & $\delta_{sys}P$ & $\delta_{total}P$ & $\delta_{\text{scl}}P/P$  \\ \hline
3.47	& $-$1 to 1	& $-$0.011	& 0.12	& 0.022	& 0.12 & 0.026 \\
4.09	& $-$1 to 1	& $-$0.089	& 0.12	& 0.022	& 0.12 & 0.026\\
4.88 & $-$1 to 1	& 0.006	& 0.13	& 0.022	& 0.13 & 0.026\\ \hline
2.8 to 5.5 	& $-$0.79	& $-$0.045	& 0.12	& 0.022	& 0.12 & 0.026\\
2.8 to 5.5 	& $-$0.41	& 0.15	& 0.12	& 0.022	& 0.12 & 0.026\\
2.8 to 5.5 	& 0.19	& $-$0.19	& 0.12	& 0.022	& 0.12 & 0.026\\   \hline

3.47						& $-$0.80 	& $-$0.088	& 0.21	& 0.022	& 0.21 & 0.026\\
4.10						& $-$0.79		& $-$0.14	& 0.20	& 0.022	& 0.20 & 0.026\\
4.86						& $-$0.77		& 0.036	& 0.22	& 0.022	& 0.22 & 0.026\\

3.45						& $-$0.44 	& 0.15	& 0.20	& 0.022	& 0.20 & 0.026\\
4.09						& $-$0.40 	& 0.16	& 0.22	& 0.022	& 0.22 & 0.026\\
4.88						& $-$0.36 	& 0.10	& 0.22	& 0.022	& 0.22 & 0.026\\

3.50						& 0.12	 	& $-$0.10	& 0.20	& 0.022	& 0.20 &  0.026\\
4.10						& 0.19		& $-$0.27	& 0.21	& 0.022	& 0.21 & 0.026 \\
4.90						& 0.26		&  $-$0.12	& 0.21	& 0.022	& 0.22 & 0.026 \\ \hline
\end{tabular}}
\caption{Summary of $P$ measurements and uncertainties. The values of $E_\gamma$ and $\cos\theta_\Xi$ given are the means of their distributions within each bin.}
\label{tab:P}

\end{table}
\begin{table}[tbh]
\center
\resizebox{0.49\textwidth}{!}{\begin{tabular}{c c c c c c c }

\hline\hline
$E_\gamma$ (GeV) & $\cos\theta_\Xi$ & $C_x$ & $\delta_{stat}C$ & $\delta_{sys}C$ & $\delta_{total}C$ & $\delta_{scl}C/C$  \\ \hline
3.47	& $-$1 to 1		&  0.21	& 0.39	& 0.01	& 0.39 & 0.039\\
4.09	& $-$1 to 1		& $-$0.083	& 0.34	& 0.01	& 0.34 & 0.039\\
4.88	& $-$1 to 1		& $-$0.021	& 0.32	& 0.01	& 0.32  & 0.039\\ \hline
2.8 to 5.5		& $-$0.79		& $-$0.21	& 0.33	& 0.01	& 0.33 & 0.039\\
2.8 to 5.5 		& $-$0.41		& 0.37	& 0.35	& 0.01 	& 0.40 & 0.039\\
2.8 to 5.5 		& 0.19		& 0.012	& 0.40	& 0.01	& 0.40 & 0.039 \\ \hline
$E_\gamma$ (GeV) & $\cos\theta_\Xi$ & $C_z$ & $\delta_{stat}C$ & $\delta_{sys}C$ & $\delta_{total}C$ & $\delta_{scl}C/C$  \\ \hline
3.47	& $-$1 to 1		& 0.52	& 0.35	& 0.052	& 0.35 & 0.039\\
4.09	& $-$1 to 1		& 0.67	& 0.29	& 0.052	& 0.29 & 0.039\\
4.88 	& $-$1 to 1		& 0.001	& 0.26	& 0.052	& 0.26 & 0.039\\ \hline
2.8 to 5.5 		& $-$0.79		& 0.52	& 0.32	& 0.052	& 0.33 & 0.039\\
2.8 to 5.5 		& $-$0.41		& 0.49	& 0.28	& 0.052 	& 0.29 & 0.039\\
2.8 to 5.5		& 0.19		& 0.13	& 0.30	& 0.052	& 0.30 & 0.039\\ \hline
\end{tabular}}
\caption{Summary of $C_x$ and $C_z$ measurements and uncertainties.}
\label{tab:Cxz}
\end{table}

For comparison, the polarization predictions of the three phenomenological model variants put forth by Refs.~\cite{Nakayama:2006ty, Man:2011np} to help explain the differential cross 
sections reported by Ref.~\cite{Guo:2007dw}, overlay our results in Figs.~\ref{fig:PCxCzE}, \ref{fig:PCxCzXi}, and \ref{fig:P9Xi}.  All three model variants share the same framework, in which cascade photoproduction proceeds through the decay of intermediate hyperon resonances that are produced via relativistic meson exchange. The predictions are based on pseudoscalar (solid red) and pseudovector (dashed blue) relativistic meson-exchange. Contributions from the $\Sigma$(2030), which has spin-7/2, were introduced in Ref.~\cite{Man:2011np} (dotted green).

 The predicted values of $P$ and $C_x$ follow fairly flat curves, that when 
determined over the entire angular and/or energy range, integrate to nearly zero. Conversely,  the predicted values of $C_z$ are positive and sizable over the kinematic range and thus do not integrate to zero on any interval.  

As shown in Figs.~\ref{fig:PCxCzE}, \ref{fig:PCxCzXi}, and \ref{fig:P9Xi}, our measurements are generally well described by the pseudoscalar (solid red) and the 2011 pseudovector (dotted green) models but not the 2006 pseudovector model (dashed blue). We have performed a statistical comparison of the three model variants to 15 independent data points, 9 of which come from the induced polarization, $P$, in the un-integrated binning scheme in Table~\ref{tab:P}, while the other 6 data points come from the transferred polarization, $C_x$ and $C_z$, summed over $E_\gamma$. The agreement between the data and the pseudoscalar variant is good, with a $\chi^2=13.0$. The 2006 variant of the pseudovector model has $\chi^2=33.0$ and is therefore excluded by the data with $\sim99\%$ confidence. The 2011 variant of the pseudoscalar model (dotted green) has $\chi^2=17.4$. Similar results are found when comparing the model to the $\cos\theta_\Xi$ integrated transferred polarization results. However it is import to point out these models were tested against the cross sections measurements up to  around $4$~GeV. Above that, it is possible that other mechanisms not accounted for such as the Regge trajectories and other higher-mass hyperons might need to be included.


 Finally, it is worth pointing out that the photoproduced $\Lambda$ was 
 observed~\cite{Bradford:2006ba} to exhibit nearly $100\%$ polarization by evaluation of $R=\sqrt{C_x^2+C_z^2+P^2}$. This quantity for the $\Xi^-$, integrating our results over all bins, is 
 $0.30\pm0.14$, which is non-zero but significantly smaller than the $\Lambda$ counterpart. 

\section{Conclusion}
To summarize, we have made the first polarization measurements for the $\Xi^-$ in photoproduction by measuring the induced polarization, $P$, as well as transferred polarization, $C_x$ and $C_z$, using a circularly polarized photon beam.  We have found that the total integrated $\Xi^-$ polarization departs from zero by 2$\sigma$, but is significantly smaller than in the analogous case for $\Lambda$ photoproduction. The results have been compared, and show general agreement with the predictions of a phenomenological model of cascade photoproduction involving intermediate hyperon resonances that are produced, predominantly in the $t$-channel, via relativistic pseudoscalar meson exchange. The results strongly disfavored a model variant that includes significant contributions from the $\Sigma(2030) \frac{7}{2}^+$. While precisely distinguishing between current and future model variants to determine the role of high-spin excited hyperons and the contributions from scalar versus vector exchange mechanisms will be left to future experiments at CLAS12 and GlueX~\cite{gluex}, we have made the first step toward a detailed understanding of $\Xi^-$ photoproduction.

  


\section{Acknowledgments}
We thank K.~Nakayama for many fruitful discussions in which he provided his insight and support. We acknowledge the outstanding efforts of the staff of the Accelerator and the Physics Divisions at Jefferson Lab that made 
this experiment possible. This work was supported in part by the U.S. Department of Energy, the National Science Foundation, the Italian Istituto Nazionale di Fisica Nucleare, the 
French Centre National de la Recherche Scientifique, the French Commissariat \`{a} l'Energie Atomique, the National Research Foundation of Korea, the UK Science and
Technology Facilities Council (STFC), and the Physics Department at Moscow State University. The Jefferson Science Associates (JSA) operates the Thomas Jefferson
National Accelerator Facility for the United States Department of Energy under contract DE-AC05-06OR23177. The FIU group is supported by  the U.S. Department of Energy, Office 
of Nuclear Physics, under contracts No.~DE-SC0013620.

\section{Bibliography}
\bibliographystyle{elsarticle-num}
\bibliography{XiPol}


\end{document}